# Low-energy neutron scattering on light nuclei and $^{19}$B as a $^{17}$B-$n$-$n$ three-body system in the unitary limit

Jaume Carbonell[1⋆], Emiko Hiyama[2,3], Rimantas Lazauskas[4], and F. Miguel Marqués[5]

**1** Institut de Physique Nucléaire, Univ. Paris-Sud, IN2P3-CNRS, 91406 Orsay Cedex, France
**2** Department of Physics, Kyushu University, Fukuoka, 819-0395, Japan
**3** RIKEN Nishina Center, Wako 351-0198 Japan
**4** IPHC, IN2P3-CNRS/Université Louis Pasteur BP 28, F-67037 Strasbourg Cedex 2, France
**5** LPC Caen, Univ. Normandie, ENSICAEN, Univ. Caen, CNRS/IN2P3, 14050 Caen, France
⋆ carbonell@ipno.in2p3.fr

## Abstract

**We consider the evolution of the neutron-nucleus scattering length for the lightest nuclei. We show that, when increasing the number of neutrons in the target nucleus, the strong Pauli repulsion is weakened and the balance with the attractive nucleon-nucleon interaction results into a resonant virtual state in $^{18}$B. We describe $^{19}$B in terms of a $^{17}$B-$n$-$n$ three-body system where the two-body subsystems $^{17}$B-$n$ and $n$-$n$ are unbound (virtual) states close to the unitary limit. The energy of $^{19}$B ground state is well reproduced and two low-lying resonances are predicted. Their eventual link with the Efimov physics is discussed. This model can be extended to describe the recently discovered resonant states in $^{20,21}$B.**

## Contents

# 1 Introduction

One of the most interesting things one can study in experimental and theoretical nuclear physics is the very low energy (S-wave) scattering of neutrons ($n$) on a nuclear target ($A$). Free from the Coulomb repulsion, centrifugal barrier, spin-orbit and tensor (diagonal part) terms as well as from the spurious kinetic energy, this process is very sensitive to the effects of the strong interaction in its simplest expression. When a low energy $n$ hits on a heavy nucleus, it gives rise to fantastic forest of resonances, as the one illustrated in left panel of Figure 1 and taken from [1, 2], obtained in a n-$^{241}$Am scattering experiment performed at n—TOF CERN facility [3, 4]. A series of S- P- and D- waves resonances at the energy scale of eV, i.e. million times smaller than the typical nuclear energies, are nowadays well identified and some of them with the assigned $J^{\pi}$ quantum numbers (see [5] for a recent review). This very low energy $n$ collisions can be quite catastrophic as well, since a few meV neutron is enough to broke an Uranius nucleus into pieces with the the well known enriching or devasting consequences..

With the light nuclei the situation is less dramatic but it has its own interest. We displayed in the right panel of Figure 1 the first resonance in nuclear physics, n-$^{3}$H P-wave at $E_{cm} \approx 3$ MeV, which made the delicious of theorists since it was shown to be very challenging for the nuclear interactions (see e.g. Refs. [6, 7]). Inspite of being less spectacular than for heavy nuclei, we will see in what follows that the systematic study of nA scattering by increasing complexity give rise to interesting phenomena allowing a very simple description of nuclear dynamics.

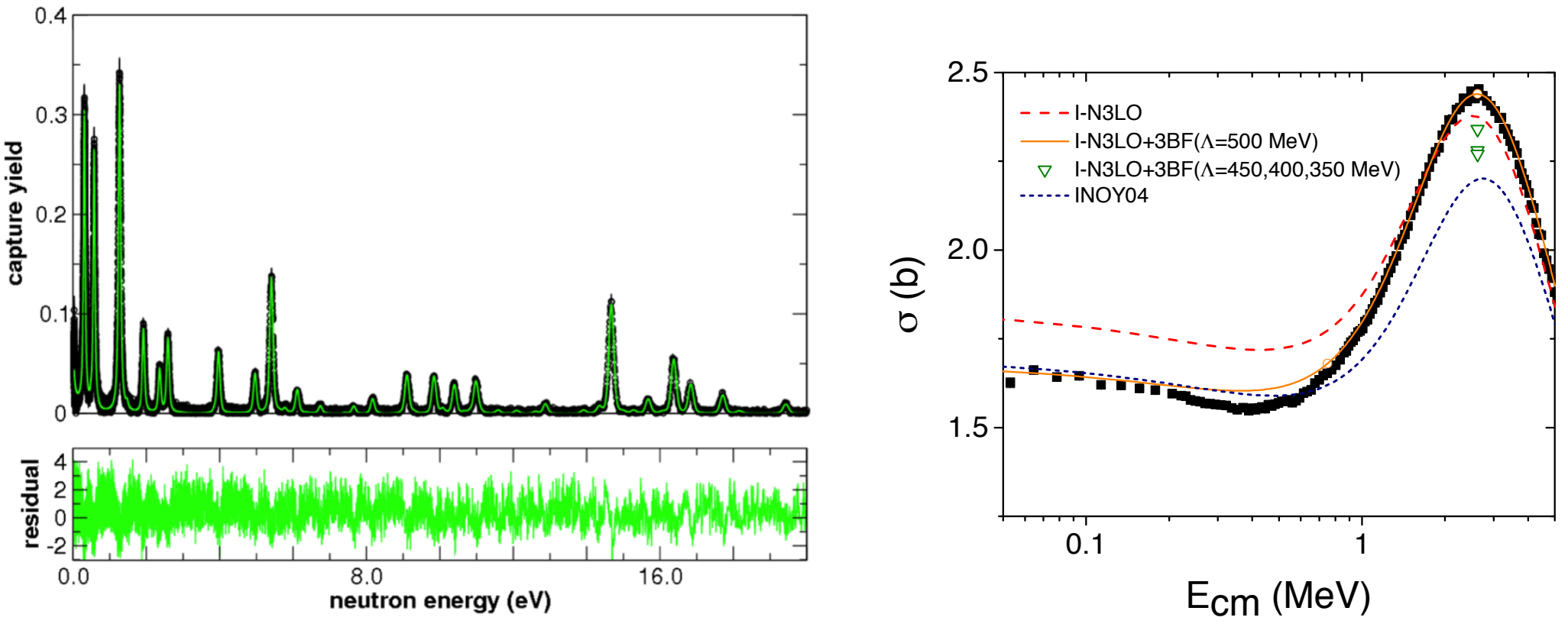

Figure 1: Left panel: low energy cross section of n on $^{241}$Am displaying a dense ensemble of S-waves resonances at the eV energy range (from [1, 2]). Right panel: n-$^{3}$H low energy cross section displaying a resonant behaviour at $E_{cm} \approx 3$ MeV

# 2 $n$-A scattering length

The S-wave neutron-Nucleon interaction ($n$N) is globally attractive in all the spin and isospin channels. In the $np$ case, the triplet state ($^{3}S_{1}$) is sensibly more attractive than the singlet one ($^{1}S_{0}$) and – together with the tensor coupling – is able to generate the fist nuclear structure: the deuteron bound state. In the singlet one the $np$ system remains a virtual state although very close to threshold. In the $nn$ case, the $^{1}S_{0}$ potential is very close to the $np$ one and has also a nearthreshold virtual state. These potentials are shown in figure 2 (left panel) in a simple interaction model, together with the poles of the corresponding scattering amplitudes in the complex momentum plane (right panel). The spin-dependence of $V_{nN}$ manifests in a 20% variation in the attractive strength of the $^3S_1$ versus $^1S_0$ potentials.

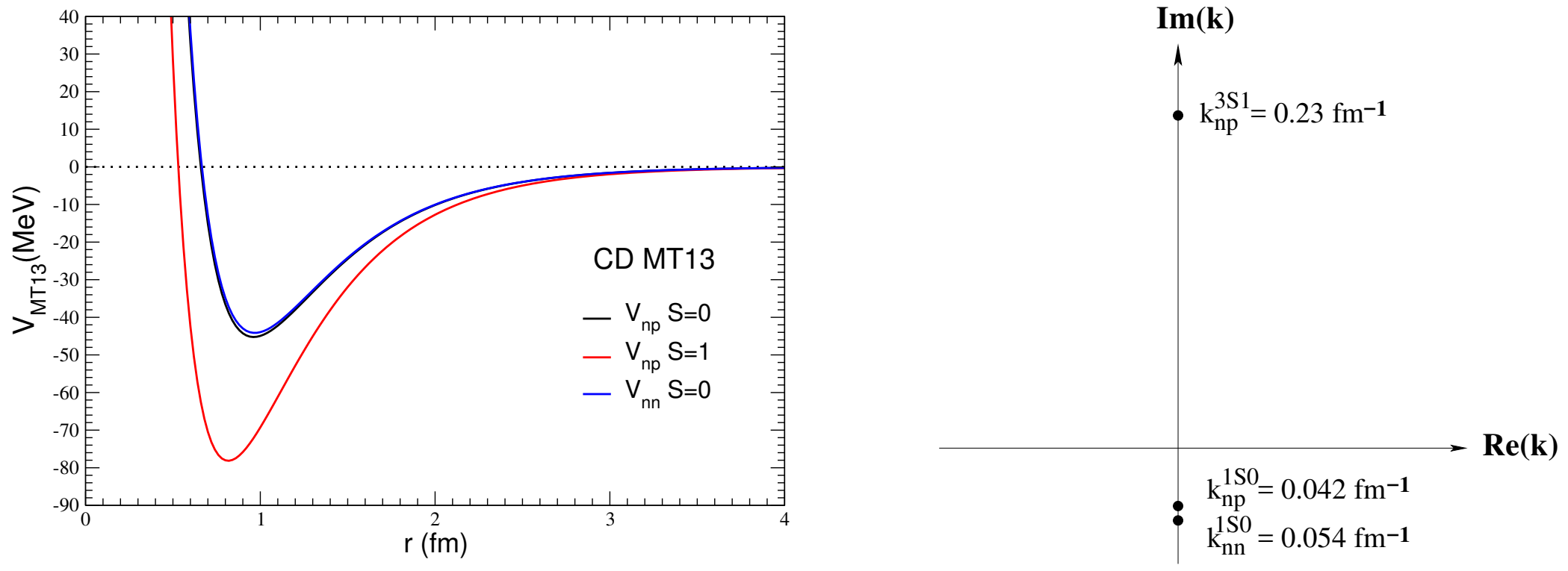

Figure 2: S-wave $V_{nN}$ in the different spin an isospin channels (left) and singularities of the corresponding scattering amplitude in the complex momentum plane (right).

Despite all $V_{nN}$ are attractive, a low energy $n$ scattering on a nucleus (A) feels the other n in the target and will soon behave – in fact starting by deuteron – as if the $n$A potential $V_{nA}$ were repulsive. The Pauli principle – imposing an antisymmetric wave function – acts *as if* there was a repulsive interaction among $n$'s. This has dramatic consequences in the 3$n$ and 4$n$ systems: the 3$n$ Hamiltonian, has a ground state bound by $\approx$ 1 MeV but this state is symmetric in particle exchange and not realised in Nature. Thinking in terms of Harmonic Oscillator discret basis on each Jacobi coordinates [8], the first antisymmetric (A) solution is several tens of MeV above the symmetric (S) one (see Figure 3) with several mixed-symmetry (MS) states in between. This is why the 3$n$ system is not only unbound but it requires an unphysically large enhancement of the interaction to be observed as a resonance [9–11]. The same happens with the 4$n$ hamiltonian [10–12] with a symmetric ground state bound by $\approx$ 5 MeV. As we will see in what follows, Pauli principle plays also a relevant role in understanding the main features of the low energy scattering of $n$'s on light nuclei.

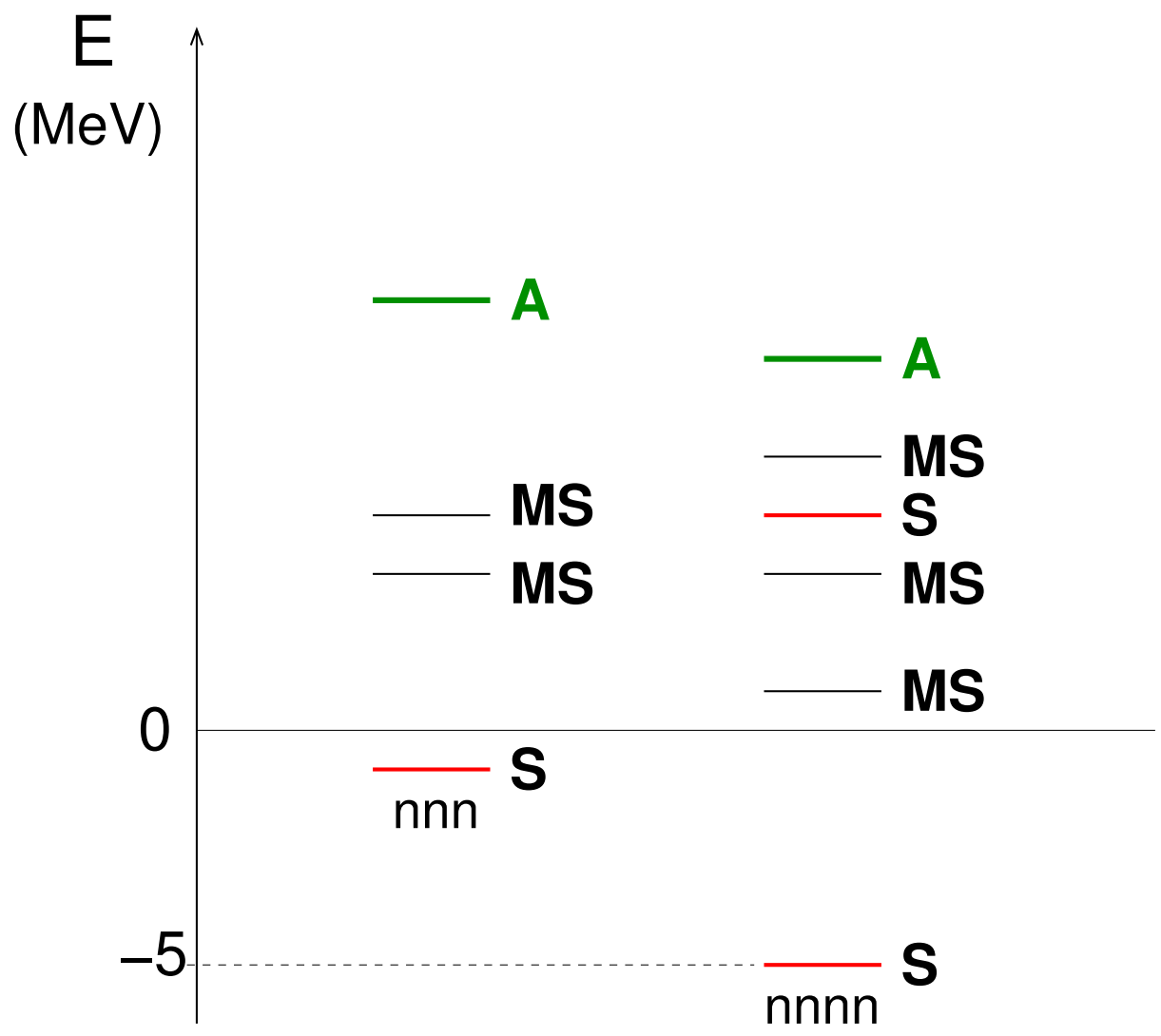

Figure 3: Schematic low level ordering of 3n and 4n in OH basis displaying the first symmetric (S) antisymmetric (A) and mixed-symmetry (MS) excitations.

A pertinent observable quantity to measure the repulsive or attractive character of an in-teraction is the scattering length, which can be defined as the $n$A scattering amplitude at zero energy $a_s = -f_{nA}(E=0)$. For a purely repulsive potential this quantity is always positive. For a purely attractive one, its sign depends on the strength of the potential: it starts being negative for a weak potential but changes its sign after the first bound state appears. This behaviour is generically illustrated in figure 4 in the case of a square well potential. In a more realistic interaction – mixing repulsive core with attractive long range part – the sign results from a balance among both tendencies.

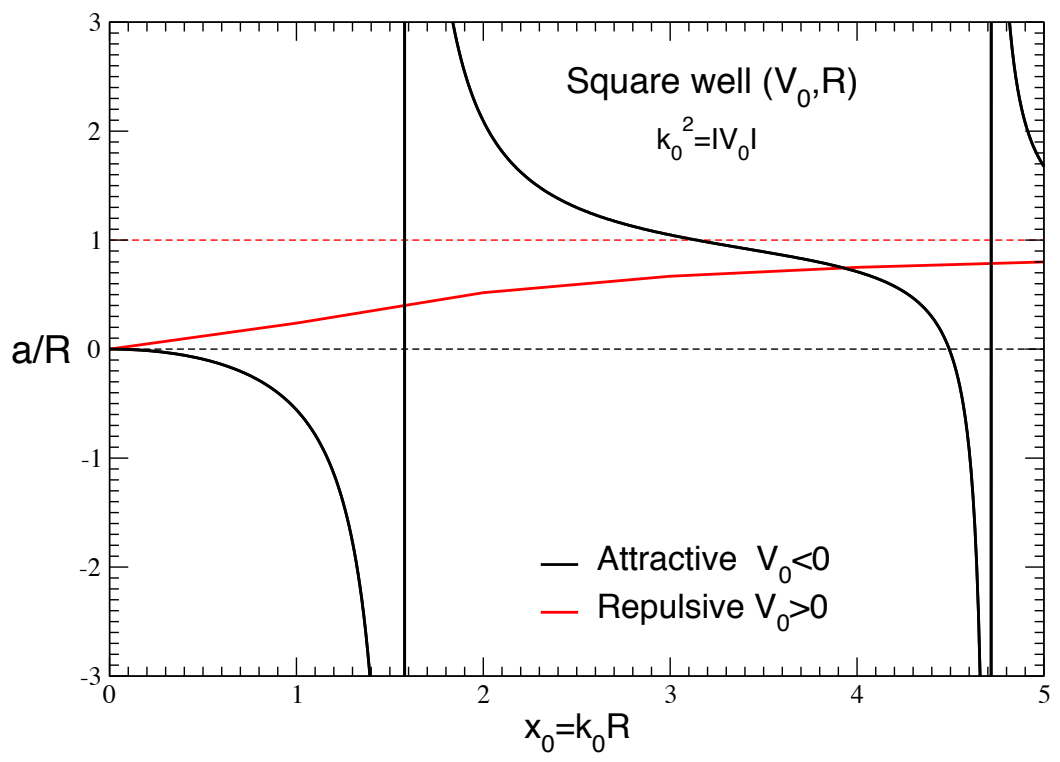

Figure 4: Scattering length in a square well potential with parameters $(V_0, R)$ as a function of its strength parameter $x_0 = \sqrt{V_0}R$. Solid red curve correspond to the repulsive case and solid black one to the attractive case. The later displays a pole singularity at the strength values $x_0$ where a new bound states appears in the system, e.g. $x_0 \approx 1.58$ and $x_0 \approx 4.72$ respectively.

We have displayed in Table 1 the evolution of the $n$A scattering length $a_{nA}$ when increasing the number of neutrons (N) in the nucleus target. We used the notation $a_\pm$ to denote the total spin $S{=}J_A \pm 1/2$, coupling the $n$ and the target on $J_A$ spins, keeping the value $a_-$ for the case $J{=}0$. Most of the experimental values in Table 1 are taken from the compilations [13, 14]. In some cases, where incompatible results are assigned to the same reaction, like for instance for the n-$^3$H, we have chosen the "Recommended value" or our personal conclusion guided by some theoretical input (see [6, 15, 16] for a discussion). In other cases, the quoted value results from a theoretical analysis of experimental data [19, 20].

With **N=0** (A=$p$) the two $np$ scattering lengths are attractive, as expected from figure 2. The positive value of $a_+$ in the np S=1 channel (denoted with an asterisk +5.42*) is due to the existence of the $^2$H bound state (deuteron) and so to an attractive channel despite its sign, as illustrated in figure 4.

Apart from the Pauli forbidden $nn$ $^3S_1$ state, the first repulsive channel appears already with **N=1**, in the S=3/2 $n$-$^2$H and in the S=1 $n$-$^3$He maximal spin states (corresponding to $a_+$). In the (schematic) shell model representation of the compound system, the two neutron spins are aligned and occupy s-wave orbitals, what is forbidden by Pauli principle. In nature, this schematic representation manifest as a repulsive state. Notice that the $a_-$ values of these systems, corresponding to antilaligned neutron spins (that is the S=1/2 $n$+$^2$H state and the S=0 $n$+$^3$He one), are not affected by Pauli and are naturally attractive in both cases despite its positive value corresponding to triton $^3$H ($a_-$=0.65* fm) and $^4$He ($a_-$=6.6*-3.7i) bound state. In the later case, there is a negative imaginary part due to its coupling to the $p$-$^3$H channel.

With **N=2** ($n$-$^3$H and $n$-$^4$He) all the scattering lengths are repulsive. The same happens for the unique **N=3** state: n$^6$Li. It si worth noticing that this strong repulsion – manifested from $^2$H to $^6$Li – manifest only in S-wave. Most of these systems have attractive nA P-waves, which in some case like in n-$^3$H displayed in Figure 1 (see [21]) and $^4$He [22–24] are even resonant.

| N Z A | Symbol | $J_A^\pi$ | $a_-$ | $a_+$ | Ref. |
|---|---|---|---|---|---|
| 0 1 1 | p | $1/2^+$ | -23.71 | +5.42* | [13,14] |
| 1 0 1 | n | $1/2^+$ | -18.59 | ⊘ | [13,14] |
| 1 2 | $^2$H | $1^-$ | +0.65* | +6.35 | [13,14] |
| 2 3 | $^3$He | $1/2^+$ | +6.6* -3.7i | +3.5 | [13,14] |
| 2 1 3 | $^3$H | $1/2^+$ | +3.9 | +3.6 | [16] |
| 2 4 | $^4$He | $0^+$ | +2.61 | | [13,14] |
| 3 3 6 | $^6$Li | $1^+$ | +4.0 | +0.57 | [13,14] |
| 4 3 7 | $^7$Li | $3/2^-$ | +0.87 | -3.63 | [13,14] |
| 6 2 8 | $^8$He | $0^+$ | -3.17 | | [17,18] |
| 3 9 | $^9$Li | $3/2^-$ | ≈-14 | | [18] |

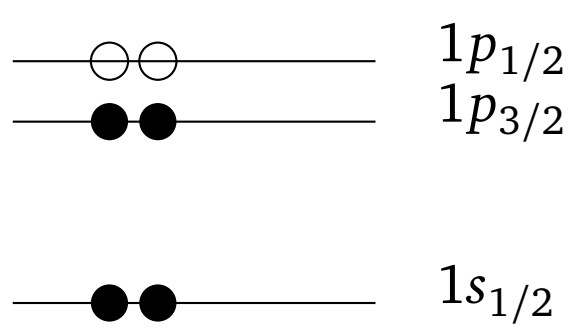

Table 1: In the left panel the spin-dependent experimental nA scattering length (fm) for light nuclei: we denote by $a_\pm$ the S=$J_A$±1/2 total spin state, keeping $a_-$ for the $J_A$=0 case or unassigned values of S (like $^9$Li). Right panel represents the filled shell model neutron orbitals in $^7$Li.

For **N=4**, the $n$-$^7$Li scattering length becomes again attractive. Two of the four $n$'s in the target fill P-wave orbital, as illustrated in the right panel of Table 1. As a consequence the effects due to antisymmetrization with the S-wave incoming $n$ is weakened and the balance between the attractive $nN$ interaction and the Pauli repulsion results in favour of an attractive S-wave state. This attraction persists in $^{12}$Be [19] and $^{15}$B [20].

When arriving at the $^{17}$B, a particle stable nuclei with groud stae $J^\pi$=$3/2^-$, the balance between the attraction and repulsion is so fine-tuned that the scattering length become huge, indicating the presence of $^{17}$B virtual state extremely close to the $n$-$^{17}$B threshold. This was first found in the MSU experiment [25], where the best fit to the data provided an $n$-$^{17}$B scattering length of $a_S = -100$ fm, which constitutes the absolute record of the whole nuclear chart [14]. However a $\chi^2$ analysis of their results allow them to fix only an upper limit $a_S < -50$ fm. Since the ground state of $^{17}$B is a $J^\pi$=$3/2^-$ state the total S of the $n$-$^{17}$B state can be $S$=$1^-$ or $S$=$2^-$ and some shell model calculation let them conclude that the measured scattering length corresponded to $S$=$2^-$. The MSU results have been recently confirmed at RIKEN [26].

This remarkable experimental finding of an extremely resonant $n$-$^{17}$B system, and the fact that an *ab initio* calculation for A=18 was beyond our scope, motivated us to model the S-wave $n$-$^{17}$B interaction and attempt to describe $^{19}$B – also experimentally known – in terms of a 3-body double resonant $^{17}$B-$n$-$n$ cluster. The main results, published in [27], are summarized in the following section.

## 3 Modeling the $n$-$^{17}$B interaction

We have modelled the resonant S-wave $n$-$^{17}$B interaction by a sum of an attractive term to account for all the $V_{nN_i}$ attraction and a repulsive one to account for the Pauli repulsion among the incoming and target n's. We have assumed the following form:

$$V_{n^{17}B}(r) \;=\; V_r\left(e^{-\mu r}-e^{-\mu R}\right)\frac{e^{-\mu r}}{r} \tag{1}$$

where $R$ is a hard-core radius, fixing the penetration of the incoming neutron in the target nucleus, and $\mu$ is a range parameter for the folded $n$-$^{17}$B potential. We fixed $R=3$ fm, which corresponds to the r.m.s. matter radius of $^{17}$B [28], and we have taken $\mu=0.7$ fm$^{-1}$ corresponding to the pion mass. At this level we work under the hypothesis of spin-independent $n$-$^{17}$B interaction, in particular the (unlikely) fact that both scattering length are equal $a_{1^-}=a_{2^-}\equiv a_S$. In this case, the only remaining free parameter is the strength $V_r$ which is adjusted to reproduce $a_S$ and so the virtual state. Since the precise value of $a_S$ is not known we have considered a wide range of variation. The dependence of $a_S$ on $V_r$, together with the corresponding potential for $a_S=-50,-100,-150$ fm, are displayed in figure 5.

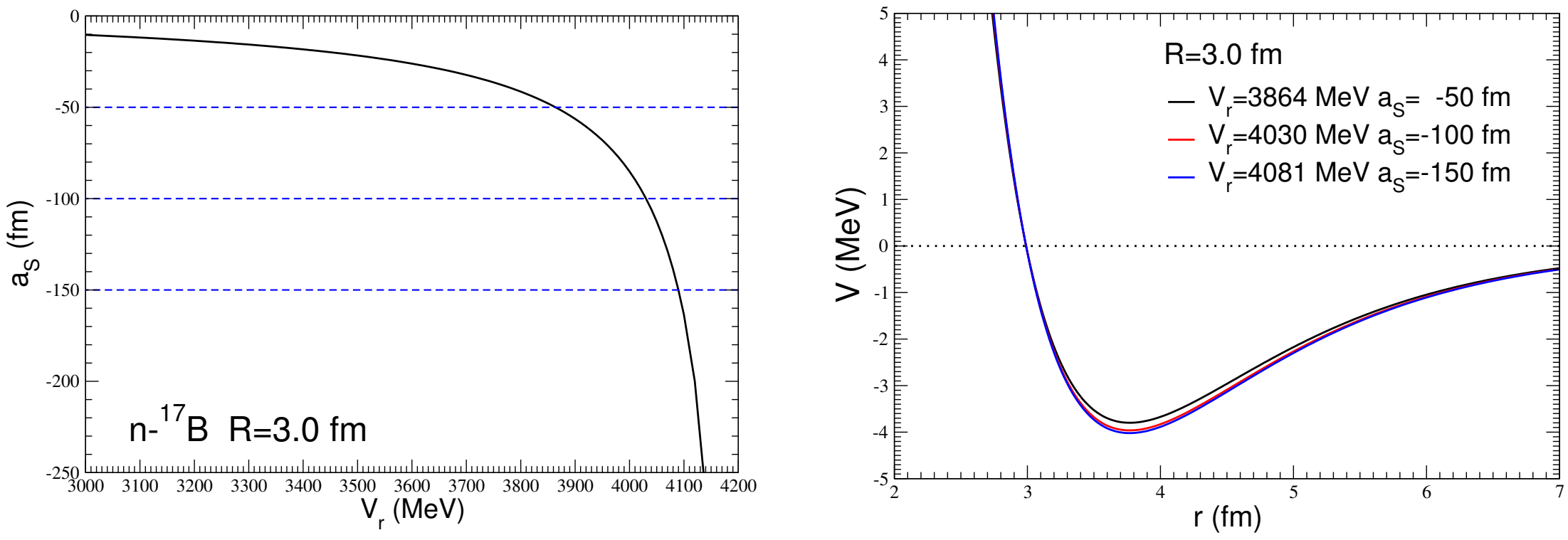

Figure 5: Dependence of the $n$-$^{17}$B scattering length $a_S$ on $V_r$ for $R=3$ fm (left panel). Dashed lines correspond to some selected values of $a_S$=-50,-100,-150 fm and corresponding strength values $V_r$ (in MeV) for which $V_{n^{17}B}$ have been drawn (right panel) .

The $^{18}$B virtual sate is one of the most fascinating systems in Nuclear Physics. We have represented in left panel of Figure 6 (black color) the n-$^{17}$B S-wave cross section $\sigma_0$ as a function of the center of mass energy E below 100 keV obtained by the above described model by assuming the value of $a_S$=-50 fm. Even in this very conservative case, a low energy $n$ scattering on $^{17}$B will feel a "monster" of geometrical size $D\sim 400$ fm (notice that $\sigma(E=0)=4\pi a_s^2$) although in a very limited energy range below few tens of MeV. The n-$^{17}$B cross section is compared to the np one in the triplet (S=1) state (red color) which appears at this energy scale totally flat. Its resonant character is however clearly manifested in the right panel when it is compared to the normal values of the nA (total) cross sections with neighbouring nuclei.

The possibility of similar strongly resonant structures, but bound instead of virtual, in the neutron rich nuclei can not be excluded. This will manifest as $nA$ states with huge and positive scattering length and so corresponding to extremely large (A+1) bound nuclei, . Such a subtle nuclear structures – only accessible via scattering experiments – could offer the possibility to "visualize" a nucleus using microscopic techniques as it is currently done with atoms.

## 4 Describing $^{19}$B as a $^{17}$B-$n$-$n$ three-body cluster

In order to describe $^{19}$B as a $^{17}$B-$n$-$n$ cluster, potential (1) has been supplemented with a realistic $n$-$n$ interaction. The three body problem was then solved using both Faddeev equations in configuration space [9] and Gaussian Expansion method [29] to inquire for the energy of the $^{19}$B ground state provided by this three-body model.

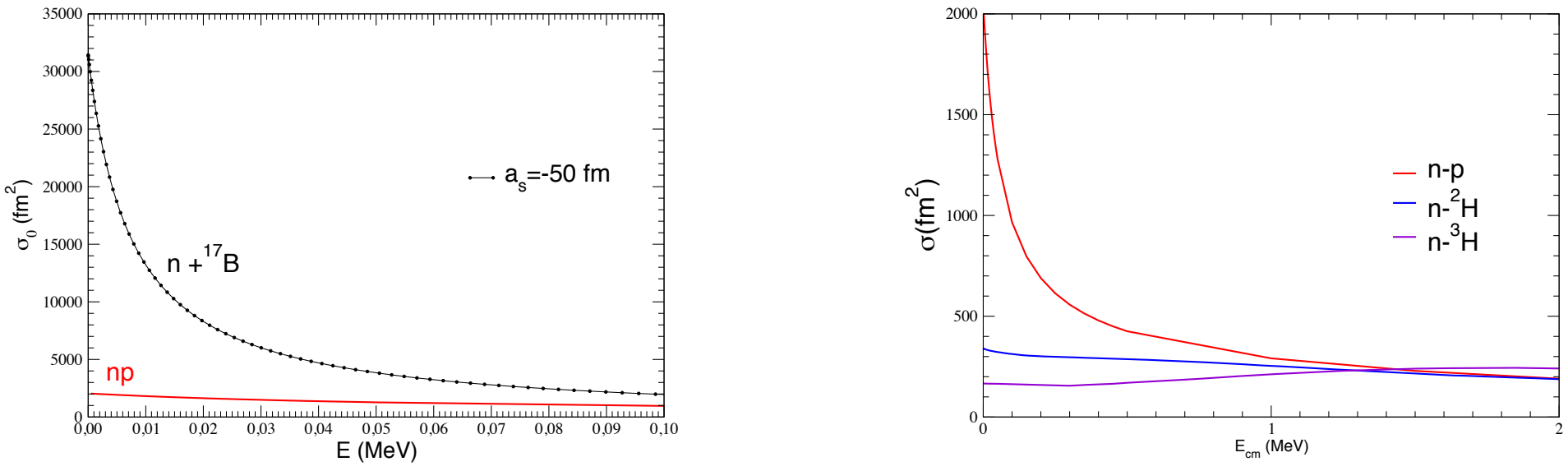

Figure 6: The low energy $n$-$^{17}$B cross section as a function of the energy for a fixed value $a_S$=-50 fm is compared to the np case (left panel), which is in its turn compared to the lightest nuclei in the rigth panel.

The results are given in figure 7 (solid blue curve) as a function of the scattering length $a_S$. $^{19}$B appears to be bound in all the range of the experimentally allowed $a_s$ values, starting from $a_s \approx$ -30 fm. The binding energy increases with $\mid a_s \mid$ and saturates at $E_u = -0.081$ MeV in the limit $a_S \to -\infty$ which is the unitary limit in the $n$-$^{17}$B channel (blue dotted points).

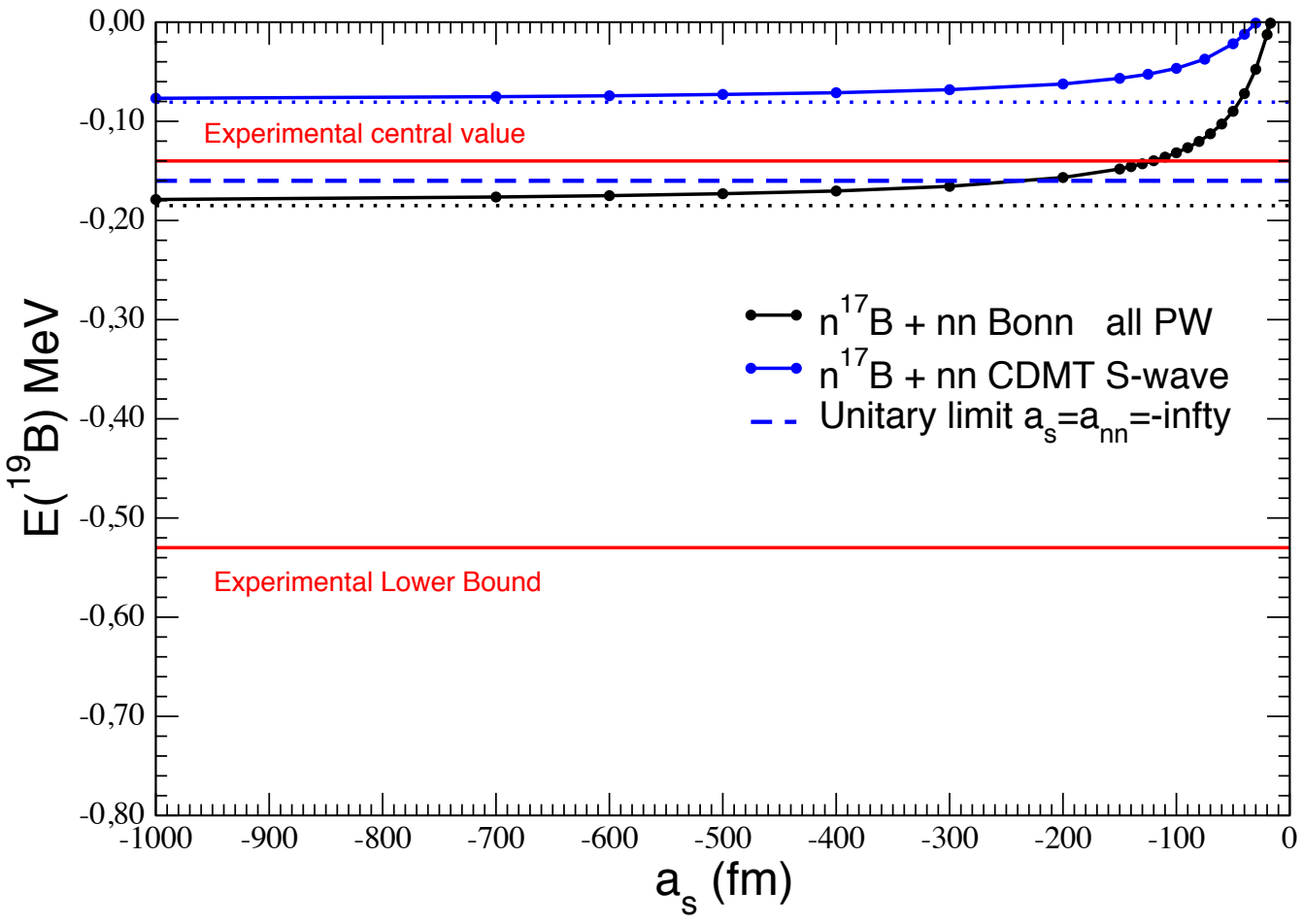

Figure 7: $^{19}$B ground-state energy with respect to the first particle threshold as a function of $a_S$, for $R = 3$ fm

These results were obtained with an S-wave $n$-$n$ interaction adjusted to reproduce the experimental $a_{nn}$=-18.59 fm (see [27] for details). In order to study the full unitary limit of the model we have also set $a_{nn} \to -\infty$, by slightly modifying the attractive part of $V_{nn}$ The results of this limit correspond to the blue dashed horizontal line $E_{uu} = -0.160$ MeV.

We have considered an alternative version of the three-body model by letting the $n$-$^{17}$B and the $n$-$n$ potentials to act in all partial waves. In this case we used the CI Bonn A model for the $n$-$n$ potential with $a_{nn} = -23.75$ fm. The results are indicated by black curves on figure 7, including the unitary limit in the given by $E_u = -0.18$ MeV. The differences come essentially from the differences in $a_{nn}$. P-wave contributions are small: for $a_S$=-150 fm they are ≈6 keV.

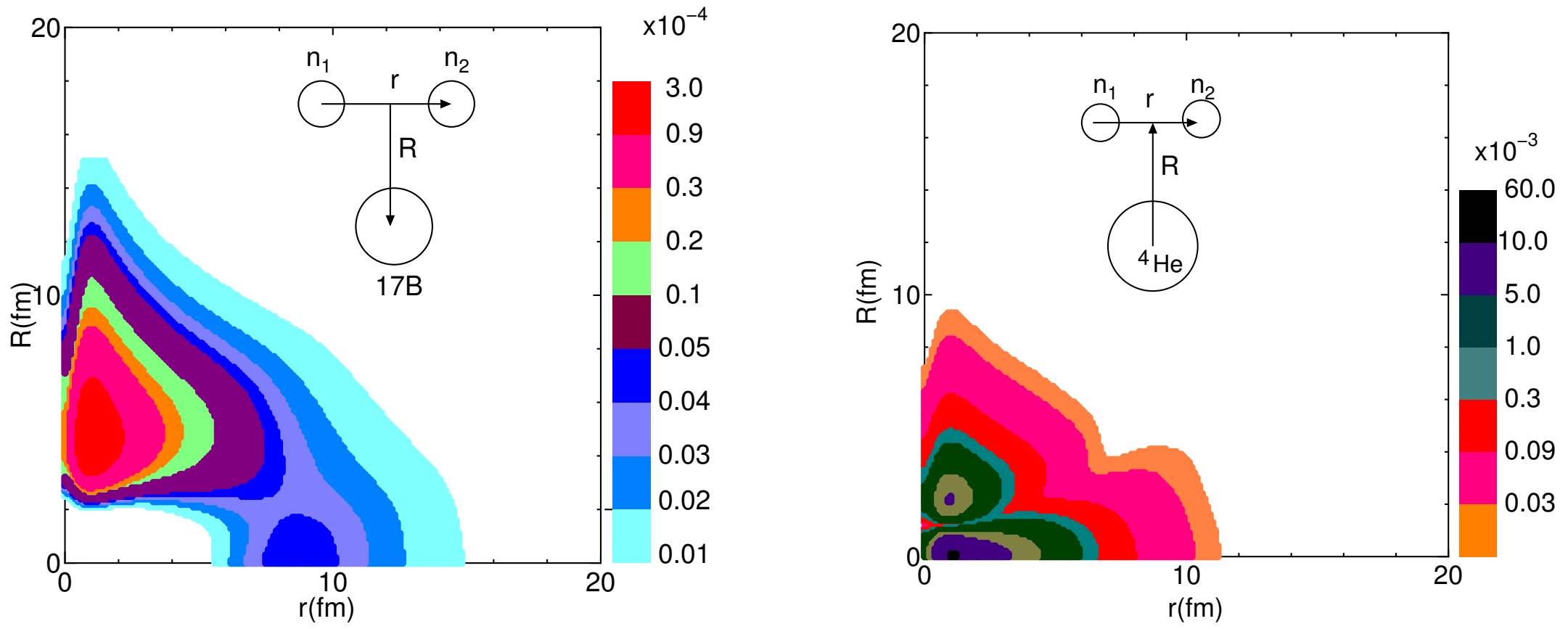

Figure 8: Ground-state probability amplitude $|\Psi(r,R)|^2$ as a function of the Jacobi coordinates: left panel $^{19}$B for $a_S = -100$ fm in model version *(ii)* and right panel for the similar structure $^6$He [30].

We would like to emphasize that due to our ignorance of the $a_S$ value we cannot predict a precise value for E($^{19}$B), but it is worth noticing that in all the domain of $a_S$ that we have considered, including the full unitary limit, the results provided by this simple model are compatible with the experimental value E=-0.14 ± 0.39 MeV.

We have represented in left panel of Figure 8 the modulus squared of the $^{19}$B ground state wave function as a function of the Jacobi coordinate $\mid \psi(r,R) \mid^2$ corresponding to $a_S$=-100 fm and $E = -0.130$ MeV. As a consequence of its weak binding energy, $^{19}$B is quite an extended dissymmetric object elongated in the $^{17}$B-($nn$) direction, as it corresponds to a two-neutron halo. Disregarding the intrinsic size of the $^{17}$B core, the calculated rms n-core distance is $R_{n-^{17}B}$ =12 fm. It is interesting to compare this result with the similar system $^6$He, considered as a $^4$He-n-n 3-body cluster (right panel of Figure 8). The corresponding value was computed in Table IV from Ref. [30] and is $R_{n-^6He}$=4.6 fm.

Apart from providing a good description of the $^{19}$B ground state, the model accommodates two broad 3-body resonances with total orbital angular momentum L=1 and L=2 respectively. Their parameters depend from $a_S$ and the model version. By fixing $a_S$=-150 fm, taking the interaction (1) in all partial waves and keeping S-wave interaction in $n$-$n$, their values are respectively $E_{L=1}$=+0.24(2)-0.31(4)i and $E_{L=2}$=+1.02(5)-1.22(6)i. Notice that the total angular momentum and parity $J^\pi$ of these resonant states results from the coupling between the quoted L and $J^\pi_{^{17}B} = 3/2^-$, which are degenerated in our calculations. Several resonances in the continuum of $^{19}$B have been observed recently, although the determination of their energies and quantum number is still in progress [26]

As mentioned at the begining of this section, the results presented above neglect any spin-spin dependence in $V_{n^{17}\mathrm{B}}$. In particular they assume the equality of the $n$-$^{17}$B scattering length in the S=1$^-$ and S=2$^-$ channels: $a_{1^-} = a_{2^-} \equiv a_S$. In our previous work [27] we have introduced a spin-dependent interaction and find the robustness of the $^{17}$B-$n$-$n$ model in what concerns the predictions of the $^{19}$B bound state: the $^{17}$B-$n$-$n$ ground state remains bound unless we introduce a spin-spin dependence at the level of $V_{n^{17}B}$ one order of magnitude greater than the one displayed in figure 2.

The large values of the $n$-$^{17}$B and $n$-$n$ scattering length, and the proximity of the $^{19}$B ground state to the unitary limit suggests that $^{19}$B could be a genuine nuclear candidate to exhibit

Efimov physics [31], that is the appearance of an infinite family of bound states with consecutive energies scaled by a universal factor $f^2$. This is indeed what would happen by setting $a_1 = a_2 = a_{nn} = -\infty$ as in the blue dashed line of Figure 7. However the $^{17}$B-$n$-$n$ system, representing a light-light-heavy structure, turns out to be quite an unfavorable case to exhibit the sequence of excited Efimov states due to the requirement of a very large factor $f$. In the real world, as well as in our model, $^{19}$B has only one bound state and it is governed by three different scattering lengths, from which only the $n$-$n$ scattering length is relatively well known and still far from $a_{nn} = -\infty$. In the case when one spin-independent $n$-$^{17}$B interaction is tuned, the universal factor turns out to be $f \approx 2000$ [31, 32]. It follows that the appearance of the first $L = 0^+$ excited state of Efimov nature in $^{19}$B would manifest only when the $n$-$^{17}$B scattering length reaches several thousands of fm. We conclude that, independently of the particular value of $a_S$, it is highly unlikely to observe any Efimov excited state in $^{19}$B. Inspite of that, it is however clear that the universal features related to Efimov physics [33] are genuinely preserved in this system. The possibility to consider the $^{19}$B ground state as being the first one of an, nonexistent, series of states can always be considered, as it was done e.g. for the triton case [31, 32]. However this is not the purpose of this work.

Finally, it is worth noticing that other attempts were made to describe the same system, like e.g. in Ref. [34], although the purely attractive (non local) $n$-$^{17}$B interaction that was chosen, overbound the $^{19}$B ground state and predicted several bound excited states. Models inspired in halo EFT [35–37] have been also used in the past to describe similar systems [38, 39]. They are based on zero-range two-body force supplemented with a three-body force required to stabilize the three-body system fixed. We have preferred to use here a more conventional approach.

## 5 Conclusion

We consider the evolution of the neutron-nucleus scattering length for the lightest nuclei. We showed that, when increasing the number of neutron in the target nucleus, the strong repulsion observed in the low energy (S-wave) neutron-nucleus interaction for light nuclei A=2-6 starts becoming attractive in $^7$Li. When filling the P- and higher angular momentum neutron orbitals in the target, the Pauli repulsion among the incoming and the target neutrons weakens and the resulting balance with the attractive nN interaction becomes globally attractive again.

In the case of $^{17}$B, this balance results into an extremely shallow virtual state which manifests by a huge scattering length $a_S$ <-50 fm, presumably in the total spin S=$2^-$ channel [25], experimentally observed but not yet precisely determined.

The possibility of similar strongly resonant structures in the nuclear chart, but being bound instead of virtual, can not be excluded. This will manifest as $nA$ states with huge and positive and scattering length and so corresponding to extremely large (A+1) bound nuclei. Such an extremely fragile nuclear structure involving sizes still smaller but close to atomic sizes – and only accessible via scattering experiments – could offer a unique possibility to "visualize" a nucleus using microscopic techniques as it is currently done with atoms.

We have constructed a simple S-wave $n$-$^{17}$B interaction model and applied it to the description of the $^{19}$B isotope as a three-body $^{19}$B-$n$-$n$ cluster state. This model describes well the energy of the $^{19}$B ground state, in agreement with the measured binding energy, and accommodates two resonant states with total orbital angular momentum L=1 and L=2, also in agreement with experimental findings [26].

The success of this simple model is to be found on the double resonant character of the interaction both in the $n$-$^{17}$B and the $n$-$n$ channels which makes the $^{19}$B nucleus a nice illustration of a system described by the unitary limit of its interactions. Despite the large values of the scattering length in each channel, the system is still far to accommodate the first Efimov excited state, due to the existence of three different scattering lengths with only one being resonant and to the asymmetry among the constituent masses would require and $a_S$ value of few thousands fm.

The proposed model can be straightforwardly extended to the description of recently observed resonant states in $^{20}$B and $^{21}$B [40] using the methods for solving the A=4 and A=5 developed in [41].

## Acknowledgements

We were granted access to the HPC resources of TGCC/IDRIS under the allocation 2018-A0030506006 made by GENCI (Grand Equipement National de Calcul Intensif). This work was supported by French IN2P3 for a theory project "Neutron-rich light unstable nuclei" and by the Japanese Grant-in-Aid for Scientific Research on Innovative Areas (No.18H05407).